\documentclass[12pt]{article}

\begin{document}

\date{}
\title{The Gibbs Paradox Revisited}
\author{Dennis Dieks \\ Institute for History and Foundations of Science \\
Utrecht University, P.O.Box 80.010 \\ 3508 TA Utrecht, The
Netherlands} \maketitle

\begin{abstract}
The Gibbs paradox has frequently been interpreted as a sign that
particles of the same kind are fundamentally indistinguishable;
and that quantum mechanics, with its identical fermions and
bosons, is indispensable for making sense of this. In this article
we shall argue, on the contrary, that analysis of the paradox
supports the idea that classical particles are always
\emph{distinguishable}. Perhaps surprisingly, this analysis
extends to quantum mechanics: even according to quantum mechanics
there can be distinguishable particles of the same kind. Our most
important general conclusion will accordingly be that the
universally accepted notion that quantum particles of the same
kind are necessarily indistinguishable rests on a confusion about
how particles are represented in quantum theory.
\end{abstract}

\section{Introduction: The Gibbs Paradox}
Imagine the following experimental set-up: a partition divides a
container into two equal parts, each containing a different ideal
gas---the amounts of gas, pressure and temperature being equal in
the two halves of the container. Now, the partition is suddenly
removed, so that the two gases start mixing via an irreversible
process; eventually a macroscopic equilibrium situation is
established. The uniform gas mixture that fills the container at
the end of this process then possesses a higher entropy than the
total entropy of the gas system we started with; the difference is
the \emph{entropy of mixing}.

The magnitude of this entropy increase can be calculated by
considering a reversible process that begins in the same unmixed
situation and ends in the same final equilibrium state as the
irreversible process we just described. The standard (theoretical)
way of realizing this reversible mixing process makes use of
semi-permeable membranes: the partition between the two halves of
the container is replaced (in thought) by two membranes, one only
transparent to the gas on the left-hand side (gas $A$, say), the
other only transparent to the other gas ($B$). These membranes can
now be slowly shifted to the left and the right wall of the
container, respectively, thereby letting gases $A$ and $B$ expand
reversibly. During this process each of the expanding gases exerts
a pressure $P$ on the membrane that is opaque to it, so work is
done. In order to keep the energy and the temperature at their
original values a compensating amount of heat, $\triangle Q$,
should therefore be supplied to the gases, and this can be
arranged by immersing the system in a heat bath. The change of
entropy resulting from this reversible process can be calculated
via $\triangle S = \int dQ/T$, with $T$ the temperature. The added
heat, $dQ$, should equal the work performed by the two gases,
i.e.\ $2 PdV$. In this way we find:
\begin{equation}\label{mixing1}
\triangle S = 2 \int PdV/T = 2 \int k N dV/V = 2 k N \log 2,
\end{equation}
where we have used the ideal gas law $PV=k N T$, with $N$ the
number of atoms or molecules in each of the two gases and $k$
Boltzmann's constant.

This entropy of mixing is independent of the exact physical
properties of gases $A$ and $B$. The only thing that plays a role
in the calculation and in the final result is that the two gases
are \emph{different}. This difference makes it possible to
design---in principle---the semi-permeable membranes that are
needed for the reversible mixing process. If the gases are the
same no distinguishing membranes can exist and there is no mixing
at all according to thermodynamics: from a thermodynamic point of
view nothing happens when the partition is removed in this case.

As a consequence, there is a discontinuity in the behavior of the
entropy: any difference between the gases, however small, produces
the same mixing entropy $2kN \log 2$, whereas there is no entropy
of mixing if the gases are the same. The existence of this
discontinuity is known as the Gibbs paradox.

Within the framework of thermodynamics the modern standard
response \cite{vankampen} to the existence of this discontinuity
is that nothing remarkable is happening here: In principle
(although not in practice) it is always possible to design
membranes that distinguish gases $A$ and $B$ as long as there is
any difference between them at all. Because there is no conceptual
difficulty in accepting a discontinuity between ``$A$ and $B$
differ'' and ``$A$ and $B$ are equal'', it should not be
considered paradoxical that there is a corresponding discontinuity
between distinguishability-in-principle plus existence of a mixing
entropy, and complete identity without such an entropy. Moreover,
in \emph{practical} situations the effectiveness of distinguishing
between two gases will be restricted by technical
limitations---this effectiveness will gradually diminish when the
two gases become more similar. As a consequence, no discontinuity
will be detectable in actual laboratory situations: the measured
mixing entropy will there vanish continuously. It is only in the
idealized situation of perfectly effective separation techniques
that the discontinuity in the entropy will manifest itself---and
as we have seen, in this case there is no conceptual problem.

\section{The Gibbs Paradox in Statistical Mechanics}

The paradox can also be formulated in statistical mechanics. In
statistical mechanics a counterpart to the thermodynamical entropy
is defined, namely the logarithm of the number of microstates $W$
that are compatible with a given macrostate: $S = k \log W$. When
an ideal gas of $N$ particles expands and doubles its volume, the
number of available microstates $X$ per particle doubles: each
particle now obtains twice as much space available to it as it had
before. This means that $W$ goes up, from $X^N$ to $(2X)^N$, which
corresponds to an entropy difference $\triangle S = kN \log2$.
When two different ideal gases mix, the statistical mechanical
entropy of mixing is therefore $2kN \log2$, exactly the value
predicted by thermodynamics.

When two equal volumes of the \emph{same} gas mix, the number of
microstates available to an arbitrary individual particle still
doubles, so the formula $S = k \log W$ gives us $\triangle S = 2kN
\log2$, as before. But now this result seems wrong, at least from
the point of view of thermodynamics. As we have seen, the
thermodynamical entropy of mixing vanishes in this case, because
nothing happens when two identical gases mix. This then leads to a
new form of the Gibbs paradox: the statistical mechanical entropy
of mixing is insensitive to the question of whether the gases are
equal or unequal, but this is in conflict with the discontinuity
predicted by thermodynamics.

In the literature the following argument is often deployed in
order to remove this discrepancy. ``Since the molecules (or atoms)
of a given species of gas are all qualitatively the same,
permutations of these particles do not have any physical effect
and do not lead to a new state; therefore, replace $W$ by $W/N!$
in the formula for the entropy.'' As it turns out, this change in
the way of counting the number of available microstates suffices
to restore agreement with thermodynamics.

Indeed, the division by $N!$ makes the mixing entropy in the
equal-gases-case disappear and leaves the other, correct results
untouched. For example, doubling of the volume of a gas without a
change in the number of particles gives us $\triangle S = kN (\log
X^N/N! - \log (2X)^N/N!)= kN\log2$, so the entropy of mixing keeps
its appropriate value in the case of two different gases that mix.
However, doubling the volume together with doubling $N$ gives us,
via Stirling's formula, that the number of microstates goes from
$W$ to $W^2$: $W = X^N/N! \rightarrow W^{\prime}=(2X)^{2N}/(2N)! =
W^2$. This implies, via $S = k \log W$, that the entropy just
doubles, without entropy of mixing, when two volumes of equal
gases are combined.

The way in which the division by $(2N)!$ in the equal-gases-case
achieves the removal of the mixing entropy is that it throws away,
in addition to the exchanges among left particles and right
particles, also all permutations in which one or more particles
coming from the left are exchanged with particles originating from
the right side of the container. This now discarded number of
permutations yields a factor $M = (2N)!/N!N!$ in the number of
microstates, which via $\triangle S =k \log M$ corresponds exactly
to the value of the entropy of mixing. In other words, there no
longer can be any mixing, nor any entropy of mixing, because the
exchange of a particle from the left with one from the right is
defined away, as not giving rise to a new physical situation.

The division by $N!$ thus restores the situation as we know it
from thermodynamics: there is a finite and constant entropy of
mixing in the case of different gases, however small the
difference between the gases may be, and there is no such entropy
in the case of equal gases.

However, this result is achieved at the price of defining away
differences that obviously are physically real, at least from the
point of view of classical physics. Indeed, classical particles
are the example \emph{par excellence} of distinguishable
individuals: no two classical particles can be in exactly the same
physical state because they will at any instant at least occupy
different spatial positions, by virtue of their impenetrability.
Moreover, classical particles follow continuous and
non-intersecting trajectories, so that they ``remember'' where
they came from. Their individuality and distinguishability are
thus preserved over time, with the consequence that it makes a
physical difference, in principle, whether a given particle in our
gas container originates from the left or from the right. So the
above resolution of the Gibbs paradox in statistical mechanics,
relying as it does on an assumed identity of states that follow
from each other by particle permutations, is in conflict with
basic features of classical mechanics. It is for this reason that
quantum mechanics is often invoked: in quantum mechanics particles
are indistinguishable as a matter of principle, so that particle
exchanges really do not alter the physical state---at least, that
is the conventional wisdom.

However, as we shall show in a moment, the starting point of the
whole argument, namely that there should be no statistical
mechanical entropy of mixing in the case of two gases of the same
kind, is shaky. Although on the macroscopic, thermodynamical level
the absence of an effect of mixing two equal gases is certainly a
justified assumption, this does not mean that there are no effects
if microscopic details are taken into account.

\section{The Statistical Mechanical Entropy of Mixing}

We are now going to argue in the context of statistical mechanics,
and this induces us to consider a slight extension of the tools
that are used in thermodynamical thought experiments. In
statistical mechanics the motivating idea is to take into account
how gases are built up from their atoms or molecules, and this
makes it natural to consider a variation on the reversible mixing
process explained in section 1. In section 1 an essential role was
played by semi-permeable membranes that were only transparent to
one of the two gases (in the case of the mixing of different
gases). In the context of thermodynamics this means that the
membranes are sensitive to chemical differences between the gases.
In statistical mechanics we have the opportunity to generalize
this and to consider membranes that are also sensitive to
microscopic particle details. In this spirit we now introduce a
new type of semi-permeable membrane: one that is transparent to
particles originating on the right-hand side of the container and
opaque to particles coming from the left-hand halve (or \emph{vice
versa}). According to classical physics such membranes are
possible in principle, as is clear from what was said above about
particle properties in classical mechanics: particles carry the
data about their origin with them, in their position and momentum
values, and this can (in principle) be used to determine whether
the membrane should stop them or not. Figuratively speaking, think
of submicroscopic computers built into the membrane that perform
an ultra-rapid calculation each time a particle hits them, to see
where it came from; or the proverbial demon with superhuman
calculational powers who stops or lets pass particles depending on
their origin. In general, of course, allowing expedients of this
kind may upset thermodynamical principles, in particular the
second law of thermodynamics. But in the thought experiment we
propose here we make a restricted use of these unusual membranes.
The idea is merely to employ them for the purpose of demonstrating
that if gases are mixed and unmixed by selection on the basis of
past particle trajectories and origins, as should be possible
according to classical mechanics, this leads to the emergence of
an entropy of mixing.

Indeed, if we use semi-permeable membranes designed in the way
just described, and employ them exactly as our earlier membranes
but now in the case of two gases of the same kind, we find just as
before that a pressure is exerted on the membranes by the
particles to which they are not transparent. Copying the reasoning
from section 1, we can conclude that this leads to the presence of
an entropy of mixing with the value $2kN \log 2$. In other words,
if the submicroscopic particle picture of statistical mechanics is
taken completely seriously, the original formula $S = k \log W$,
without the \emph{ad-hoc} division by $N!$, gives us correct
results.

In principle then, on the microscopic level of description the
mixing entropy always exists according to classical mechanics,
even in the case of equal gases. In principle, classical particles
can always be distinguished on the basis of their trajectories and
there is a corresponding mixing entropy that can be measured by
using semi-permeable membrane-like devices of the kind we have
described. Classical atoms and molecules are distinguishable
individual entities, and we can conclude that analysis of the
Gibbs paradox supports rather than undermines this general feature
of classical physics. Of course, microscopic distinguishability
cannot show up if we confine ourselves to using macroscopic
separation techniques of the kind considered in thermodynamics.
But this is a \emph{practical} matter that should not be confused
with an argument for fundamental indistinguishability on the
particle level.

\section{A New Dilemma: The Gibbs Paradox in Quantum Mechanics}

But now we are facing a new paradox. In quantum mechanics the
``identity of indistinguishable particles'' has long been
recognized as a basic principle, given concrete form by the
(anti-)symmetrization postulates. These postulates stipulate that
in the case of particles of the same kind permutations of particle
indices leave a many-particle state either invariant (the case of
bosons) or change its sign (the case of fermions); in either case
there are no measurable physical differences associated with the
permutations. These symmetrization postulates possess a law-like
validity, so they hold regardless of the peculiarities of the
situation that is considered. Therefore, from the quantum
mechanical point of view division by $N!$ seems completely
justified and even mandatory when the number of microstates has to
be determined. Application of $S = k \log W$ then seems to tell us
that as a matter of principle there can be no entropy of mixing
when two gases of the same kind mix---as we have seen above, the
division by $N!$ leads immediately to this conclusion.

This is a paradox. A treatment of the mixing of gases by means of
quantum mechanics should obviously reproduce the results of a
classical calculation in the classical limiting situation, so it
should be able to yield the value $2kN \log 2$ in the case in
which we follow individual particle trajectories, as described in
the previous section. But it now seems that according to quantum
mechanics this is impossible in principle!

To put the difficulty into perspective, consider a situation that
is possible according to quantum mechanics and at the same time
can be dealt with by classical mechanics. Suppose that the
one-particle states occurring in the quantum mechanical
many-particles wavefunction of our ideal gas system do not overlap
spatially, and that this remains true for a substantive time
interval. Well-known results (in particular Ehrenfest's theorem,
to which we shall return in section 5) assure us that in this case
the spatially isolated one-particle quantum wave packets behave
exactly like classical particles. In fact, what we have here is
the quantum mechanical description of a diluted ideal gas, and
this description is virtually identical to what classical theory
tells us: small one-particle wave packets take the place of
classical particles, are subject to the same dynamical principles,
and follow the same trajectories. This is a typical classical
limit situation, in which the predictions made by quantum
mechanics should parallel those of classical mechanics. In
particular, in the experiment with the unusual membranes of
section 3 we should expect that quantum mechanics gives us the
result we derived there, namely the existence, in principle, of an
entropy of mixing with the value $2kN \log 2$. In the limiting
situation everything goes, according to quantum mechanics, as in
the classical case and the earlier derivations can be repeated
step by step.

Apparently then, the quantum mechanical symmetrization postulates
are not decisive for the question of whether or not particles are
distinguishable in quantum mechanics! In the diluted gas situation
that we just discussed quantum particles are as distinguishable as
classical particles: they occupy different spatial positions and
follow continuous and non-intersecting trajectories. In this case
it is clear that real physical differences correspond to different
trajectories and different particle origins, and the existence of
an entropy of mixing testifies to this: the differences in
question give rise to empirical effects. Nevertheless and
paradoxically, in this very same situation the symmetrization
postulates are fully respected.

The situation becomes clearer when we consider a concrete case,
namely a two-particle quantum system in which the one-particle
wave functions do not overlap spatially, like in the diluted gas
just discussed. Take as the quantum state of the system
\begin{equation}\label{twoparticlesystem}
    |\Psi \rangle = \frac{1}{\sqrt{2}}(| \phi_1 \rangle | \psi_2
\rangle + |\psi_1 \rangle | \phi_2 \rangle ),
\end{equation}
with $|\phi \rangle$ and $|\psi \rangle$ representing two
non-overlapping wave packets. This state is symmetrical: exchange
of the indices $1$ and $2$ leaves the state invariant, and there
is thus no physical difference associated with the distinction
between these indices. Still, this symmetrical state represents a
situation in which there are two quasi-classical objects, one at
the position of the wave packet represented by $|\phi \rangle$ and
one at the position of $|\psi \rangle$. These wave packets and the
objects they represent clearly \emph{are} distinguishable, and
they are the things that are relevant for the classical limit. As
was illustrated above for the diluted gases case, these spatially
non-overlapping wave packets take the role of particles in the
classical limit.

Summarizing, although it is true that because of the
symmetrization each \emph{index} in an $N$-particles quantum state
of particles of the same kind,
\begin{equation}\label{symmetry}
    |\Psi \rangle = \frac{1}{\sqrt{N!}}\sum \Pi |\phi_{i_1}\rangle
    |\psi_{i_2} \rangle |\chi_{i_3} \rangle |\tau_{i_4}\rangle \cdots
\end{equation}---where $\Pi$ denotes permutation over the indices
and the summation is taken over all such permutations---is
associated with exactly the same ``state'' (in the sense of a
density operator obtained by partial tracing), there still will be
distinguishable particles in the classical limit if the
one-particle states $|\phi\rangle , |\psi \rangle , |\chi \rangle
, |\tau\rangle \cdots $ do not spatially overlap. Therefore, the
\emph{indices} in the quantum mechanical formalism, over which
(anti-)symmetrization takes place, cannot refer to what we
understand to be particles in classical physics!

This observation is the key to the resolution of our paradox.
Although the \emph{indices} in the ``many-particle'' quantum
formalism have completely symmetrical roles to play and do not
correspond to any physical differences, this does not entail that
it does not make a difference if we exchange two \emph{particles}
in the ordinary sense (i.e., the localized objects that we are
wont to call particles in classical physics). Therefore, there may
be a mixing entropy even if the symmetrization postulates are
fully respected: the existence of this entropy depends on the
distinguishability of particles, not on the distinguishability of
indices. The notion that the symmetrization postulates enforce
division by $N!$ in the classical expression for the entropy, and
thus make the entropy of mixing vanish for gases of the same kind,
rests on a confusion about the status of the particle concept in
quantum mechanics \cite{dieks2,lubberdinkb}.

\section{How Particles Are To Be Represented in Quantum Mechanics}

Elaborating on this conclusion, it should be noted that the
symmetrization postulates, which are responsible for the physical
equivalence of all indices in states of many particles of the same
kind, are basic postulates of quantum mechanics; they possess a
universal validity. This implies that if we were to take the
indices to refer to particles, it would follow that \emph{all}
particles of the same kind in the universe are in exactly the same
state. For example, it would not make sense to distinguish between
electrons here and electrons elsewhere, for instance in another
solar system: all electrons in the universe are ``partly here,
partly there, a bit everywhere''. The applicability of the
symmetry postulates holds regardless of what kinds of interactions
and situations are considered; in particular, whatever
circumstances should turn out to be responsible for the transition
to the classical limit, these do not affect the applicability of
the postulates. Therefore, if we were to assume that the quantum
indices refer to particles, this would imply that even in the
classical limit particles are all in exactly the same state---that
even classical particles are completely indistinguishable!

This simple reductio shows once again how wrong it is to think
that the symmetrization in quantum mechanics pertains to particles
in the ordinary sense. Classical particles are certainly
distinguishable objects, so they simply \emph{cannot} correspond
to the quantum indices in the classical limit.

Classical particles are characterized by their unique spatial
positions and trajectories. Now, as a defence of the idea that the
individuality that corresponds with these individuating features
disappears as soon as quantum mechanics is taken into account, it
is sometimes maintained that the ``haziness'' of quantum
mechanical wave functions, in the sense of their non-zero spatial
extensions, makes it impossible to follow a quantum particle in
all detail over time (e.g., \cite{cohen-tannoudji}). For this
reason precise trajectories do not exist in quantum mechanics and
the concept of genidentity, so the argument continues, cannot
apply to quantum particles: their wave packets will overlap, and
we can consequently not find out where each particle comes from
and with which earlier particle it should be considered identical.

This argument is notable for at least two reasons. First, it
apparently assumes that particles can be associated with
individual wave packets (that may overlap). This betrays a correct
intuition about what particles are and how they are to be
represented in the quantum formalism, but is of course in conflict
with the standard idea that the indices in the formalism denote
particles. Second, this ``haziness argument'' is implausible as a
general defence of the indistinguishability of particles. This
indistinguishability would apparently have to occur as soon as
there is overlap between wave packets; but haziness is something
gradual, subject to becoming more or less, and evidently not
strong enough a premiss to bear such an all-or-nothing conclusion.
Moreover, the (anti-)symmetrization postulates remain in force
even if wave packets do not overlap at all, which demonstrates
that overlap cannot be the decisive factor here.

What the haziness argument actually supports is our earlier
conclusion that particles in quantum mechanics should be
represented by one-particle wave packets, and that there is a
gradual transition from the ``classical'' situation, in which such
wave packets are spatially disjoint, to the more typical quantum
case in which there is overlap and in which the classical particle
concept is not fully applicable. Our essential argument is that
there is no relation between the particle concept thus understood
and the ``particle indices'' in the quantum mechanical formalism.

That quantum mechanics is indeed capable of representing classical
particles in the way just explained, is (as we already mentioned
earlier) guaranteed by Ehrenfest's theorem. In the case of a
Hamiltonian $H = p^2 /2m + V(r)$, with $p$ the momentum, $m$ the
particle mass and $V(r)$ a potential field, we can introduce a
force field $F(r) = -\nabla V(r)$, in terms of which Ehrenfest's
theorem takes the form
\begin{equation}\label{Ehrenfest}
\langle F(r) \rangle = m \frac{d^2}{dt^2} \langle r \rangle .
\end{equation}
For certain specific potentials (in particular free motion, i.e.\
F=0, relevant to our ideal gases case) we find that $\langle F(r)
\rangle$ equals $ F(\langle r \rangle)$, so that in these cases
the mean value of $r$ exactly satisfies the classical law of
motion $F(\langle r \rangle) = m \frac{d^2}{dt^2} \langle r
\rangle$. In general this is not so. But if the wave function is
localized in a sufficiently small region of space, so that the
variation of the force field within that region is small, we can
replace Eq.\ (\ref{Ehrenfest}) by the classical equation in a good
approximation (which becomes better when the state becomes more
localized). From this it follows that well-localized
single-particle quantum states (localized in the sense that their
associated wave packets are very narrow) approximately follow
classical trajectories. Classical trajectories thus do exist in
quantum mechanics: they are realized by (very) small wave packets.
Consequently it is essential, for the classical limit, to have a
mechanism that keeps wave packets narrow during appreciable time
intervals.

Such considerations are standard in studies on the classical limit
of quantum mechanics, and there is growing agreement that the
essential element in explaining how classical mechanics emerges
from quantum mechanics is the process of decoherence. Decoherence
processes cause the destruction of coherence between parts of the
wavefunction centered around different positions, and effectively
reduce wave functions to narrow wave packets (see for a more
extensive discussion \cite{dieks2}).

It is generally acknowledged then that the classical limit of
quantum mechanics is characterized by the emergence of classical
trajectories followed by narrow wave packets. These narrow wave
packets become the particles we are familiar with in classical
physics.

\section{Particles as \emph{Emergent} Entities}

Our conclusion is therefore that we should think of
\emph{particles}, as we know them from classical physics, as
represented in quantum mechanics by localized wave packets
\cite{dieks2,lubberdinkb}. That is to say, if we encounter a state
$|\Psi\rangle$ defined in an $n$-fold tensor product Hilbert space
$\mathcal{H}_1 \bigotimes \mathcal{H}_2 \bigotimes \mathcal{H}_3
\bigotimes ...\bigotimes \mathcal{H}_n$, and wish to investigate
whether it can be interpreted in terms of particles, we have to
ask ourselves whether it can be written as a (anti-)symmetrized
product of localized one-particle states. It is easy to show that
\emph{if} such a ``particle decomposition'' of $|\Psi\rangle$
exists, it is unique \cite{dieks2}.

In most cases states will \emph{not} allow a particle
interpretation; think, for example, of a state of the form
(\ref{twoparticlesystem}) with two \emph{overlapping} wave packets
$| \phi \rangle$ and $| \psi \rangle$ (each defined in a connected
region of space). The particle decomposition that we need, in
terms of localized states that are non-overlapping (and therefore
mutually orthogonal) clearly does not exist in this case: of
course there does exist a bi-orthogonal Schmidt decomposition, but
the states occurring in it will be linear combinations of $| \phi
\rangle$ and $| \psi \rangle$ and will consequently spatially
overlap. An arbitrarily chosen quantum state will therefore not
represent particles. We need special circumstances to make the
particle concept applicable. In this sense, the classical limit
with its decoherence processes makes classical particles
\emph{emerge} from the substrate of the quantum world.

It may be added that the circumstances that are responsible for
the emergence of classical particles at the same time justify the
use of the statistics that we expect for the case of independent
individuals. Indeed, in the case of spatially non-overlapping wave
packets, in which our particle concept becomes applicable, both
Fermi-Dirac and Bose-Einstein statistics reduce to classical
Boltzmann statistics \cite{dieks0,french}.

\section{The Gibbs Paradox: Conclusion}

When equal gases of the same pressure and temperature mix, nothing
happens from a macroscopic, thermodynamical point of view. So
there should be no entropy of mixing in this case, in conformity
with what thermodynamics predicts. In the literature this
vanishing of the thermodynamic mixing entropy when two equal gases
mix has often been interpreted as a fundamental fact, as a sign
that also on the level of statistical mechanics, when the
microscopic constitution of the gases is taken into account, it
should not make a physical difference whether gas particles
originate from one or the other of the initially separated gases.
This interpretation is mistaken. When non-thermodynamic,
microscopic separation and mixing techniques are allowed, it turns
out that even in the case of equal gases the value $2kN \log 2$ of
the mixing entropy, and other effects of mixing, can be recovered
and in principle experimentally verified. The vanishing of the
entropy of mixing is therefore conditional on looking in a purely
macroscopic way at what happens, and has no consequences for the
identity and distinguishability of microscopic particles. Invoking
quantum mechanics in order to argue that the mixing entropy
vanishes as a matter of principle, on account of the fundamental
indistinguishability of particles, is completely wrongheaded.

As it turns out, the microscopic effects of mixing classical gases
of the same kind persist in quantum mechanics. This becomes
understandable once we realize that the gas particles do not
correspond to the indices in the quantum formalism, but rather to
one-particle wave packets. In the classical limit such wave
packets become the particles we know from classical mechanics. The
conclusion that quantum particles correspond to one-particle wave
functions rather than to indices accords with other evidence that
these indices have a purely mathematical significance and do not
denote physical objects \cite{dieks,dieks1,dieks2}. According to
our analysis of what particles are, the appearance of particles
constitutes a genuine case of \emph{emergence}. Only if specific
physical conditions are satisfied, resulting in the presence of
localized wave packets (decoherence processes are usually
essential here) does the concept of a particle in the ordinary
sense become applicable to the world described by quantum
mechanics.

Finally, even in cases in which individual \emph{particles} in our
sense are not present in the mixing gases of the Gibbs paradox,
quantum mechanics predicts that a non-zero value of the entropy of
mixing can be recovered by using sophisticated membranes. The
reason is that the two initially separated volumes of ideal gas
are represented by mutually orthogonal wave functions, located on
the left and right, respectively. Since ideal gases do not
interact these wave functions remain orthogonal, and therefore
distinguishable in principle, even after the removal of the
partition. This point (to be worked out elsewhere) demonstrates
once more that the indistinguishability of bosons and fermions is
irrelevant to the resolution of the Gibbs paradox.

\end{document}